\documentclass[twocolumn]{aastex631}

\newcommand{\angstrom}{\mbox{\normalfont\AA}}
\newcommand{\magorrian}{$M_{\mathrm{BH}}$ ---$M_* \;$}

\usepackage{soul}

\usepackage{CJK}
\usepackage{amsmath}

\begin{document}

\title{Undermassive Host Galaxies of Five $z\sim6$ Luminous Quasars Detected with JWST}

\correspondingauthor{Meredith Stone}
\email{meredithstone@arizona.edu}

\author[0000-0002-9720-3255]{Meredith A. Stone}
\affiliation{Steward Observatory, University of Arizona, 933 North Cherry Avenue, Tucson, AZ 85721, USA}

\author[0000-0002-6221-1829]{Jianwei Lyu (\begin{CJK}{UTF8}{gbsn}吕建伟\end{CJK})}
\affiliation{Steward Observatory, University of Arizona,
933 North Cherry Avenue, Tucson, AZ 85721, USA}

\author[0000-0003-2303-6519]{George H. Rieke}
\affiliation{Steward Observatory, University of Arizona,
933 North Cherry Avenue, Tucson, AZ 85721, USA}

\author[0000-0002-8909-8782]{Stacey Alberts}
\affiliation{Steward Observatory, University of Arizona,
933 North Cherry Avenue, Tucson, AZ 85721, USA}

\author[0000-0003-4565-8239]{Kevin N. Hainline}
\affiliation{Steward Observatory, University of Arizona,
933 North Cherry Avenue, Tucson, AZ 85721, USA}

\begin{abstract}

We measure host galaxy stellar masses for a sample of five luminous quasars at $z\sim5-7$. Using JWST/NIRCam medium-band images of nearby PSF reference stars, we carefully subtract the contribution from the quasar light to place upper and lower limits on the flux of each host galaxy. We find that the members of our sample of quasar host galaxies have mass upper limits of $10^{9.7} - 10^{10.8} M_{\odot}$, significantly less than expected from their SMBH masses and the local \magorrian relation. We additionally obtain JWST/NIRSpec IFU spectra of three of our quasars to calculate black hole masses, which we find are consistent with those in the literature, and to search for the presence of bright but compact galaxies via a Balmer break, which we do not find evidence for. We discuss the potential effects of dust extinction on our measured fluxes and the impact of selection effects on high-redshift quasar samples. We conclude that the masses of the SMBHs relative to the host galaxy stellar masses have a much larger scatter than locally, large enough that these selection effects cannot be responsible. The result is reinforced by other studies. Finally, we explore the potential implications of these results on the picture of SMBH-galaxy coeval growth in the early Universe.

\end{abstract}


\section{Introduction} \label{sec:intro}

Active galactic nuclei (AGN) can have great effects on their host galaxies as the two evolve over cosmic time; they have the ability to enhance or halt star formation and potentially even affect the morphology of their hosts \citep[e.g.,][]{Heckman2014}. However, the precise nature of the interplay between AGN and their host galaxies is still a matter of investigation, especially at high redshift. To build a complete picture of galaxy formation and evolution and understand how AGN are triggered, it is critical to understand how supermassive black holes (SMBHs) and the galaxies in which they reside evolve together with time.

We know already that in the local Universe the stellar masses of galaxies, $M_*$---particularly their centrally concentrated bulge components---and the masses of their supermassive black holes ($M_\mathrm{BH}$) are correlated across a wide range in black hole mass. The \magorrian relation is well-studied in the local Universe and up to $z\sim3$, both in inactive galaxies \citep{Kormendy2013, Greene2020} and AGN \citep[e.g.][]{Reines2015}. At these redshifts, the host galaxies of even bright quasars are readily detectable and their stellar masses can be determined via photometry. The central point source usually outshines the host galaxy, but the host galaxy's emission can be revealed via techniques such as point-spread function (PSF) subtraction, where a theoretical or empirical PSF is scaled to the quasar image and subtracted to remove the contribution from the quasar light and reveal extended emission from the host galaxy. The black hole masses can be determined either by reverberation mapping or by the broadening of emission lines from gas orbiting near the black hole. 

Using ground-based optical telescopes and the Hubble Space Telescope, large samples of AGN (including bright quasars) at $z \lesssim 3$ have been PSF-subtracted to reveal and characterize their underlying host galaxies \citep[e.g.][]{Gehren1984, Hutchings2002, Dunlop2003, Jahnke2004, Kim2017, Yue2018, Li2021}. Above $z\sim4$, however, the rest-frame optical is shifted out of the range of these telescopes: observations instead probe the rest-frame ultraviolet, which can provide information about the star formation occurring in the host galaxy but is a less robust tracer of stellar mass. The rest-frame ultraviolet is also intrinsically fainter (and more strongly impacted by dust) in a wavelength regime where the quasar is bright. Host galaxy emission was therefore detected or constrained in only a small number of high-$z$ systems before the launch of JWST \citep[e.g.][]{Peng2006,McLeod2009,Mechtley2012,Targett2012,Schramm2019,Marshall2020}. 

Alternatively, the dynamical mass, measured via the widths of far-infrared lines like [C~II] ($158 \mu$m), can be used as an approximation of the stellar mass at high redshift. However, dynamical masses have associated uncertainties related to assumptions about the nature of the dynamics of the galaxy and its orientation on the sky. All these studies are subject to various selection biases, but their general conclusion is that up to $z \sim 4$, the local relation between stellar and SMBH mass is preserved within the uncertainties caused by these biases. That is, the black holes and stellar populations appear to co-evolve, possibly regulated by feedback processes.

JWST is advancing this topic rapidly at $z> 4$, enabling the discovery of AGN with intermediate-mass ($\sim 10^7$ M$_\odot$) black holes at $z = 4 - 11$  and increasing the number of corresponding host galaxy detections, 
finding that the SMBHs generally are overmassive relative to their host galaxies, compared with the local \magorrian relation \citep[e.g.,][]{Kocevski2023,Goulding2023, Maiolino2023, Kokorev2023}. Pushing these discoveries to very high redshifts may identify the seeds of SMBHs in the early Universe \citep{Bogdan2023,Maiolino2023}. New observations with JWST also address a related question: how seed black holes can grow to $\sim 10^9$ M$_\odot$ in $<$ 1 Gyr, as found in luminous, very high-redshift quasars \citep{Fan2023}.

We report observations addressing the latter issue using NIRCam on JWST (which probes the rest-frame optical out to much higher redshift and with greater spatial resolution and sensitivity than HST). To explore the coevolution of host galaxies and their central SMBHs at high-$z$ with NIRCam, we designed a JWST GTO program \citep[ID 1205;][]{Rieke2017} to obtain NIRCam imaging and/or NIRSpec IFU observations of five $z \sim 5-7$ quasars at a range of star formation rates, AGN luminosities, and obscurations.

Our first quasar analyzed, HSC J2239+0207 ($z=6.25$), is a sub-Eddington quasar with a high ALMA dynamical mass \citep{Izumi2019}, making its host galaxy theoretically easy to detect in PSF-subtracted NIRCam images. Indeed, we detected the host galaxy of this quasar in two bands (at signal-to-noise ratios between 5 and 7), but determined its stellar mass to be more than an order of magnitude less than its ALMA [C~II] dynamical mass \citep[][hereafter S23]{Stone2023}. J2239+0207 therefore lies significantly above the local \magorrian relation. \citet{Yue2023} searched for  host galaxies in six high redshift quasars (two of which overlap with our sample) and found similar behavior relative to the \magorrian relation. In contrast, \cite{Ding2023} detected the host galaxies of two similarly high-mass AGN and found them to be consistent with the local \magorrian relation.

Are J2239+0207 and the systems in \citet{Yue2023} outliers, coincidentally lying above the local relation and most other $z\sim6$ quasars, or does the \magorrian relation perhaps exhibit greater scatter at high-$z$ than locally? The remaining quasars in our JWST GTO program will help to expand the number of high-$M_{\mathrm{BH}}$, high-luminosity ($L_{\mathrm{AGN}}$) AGN at $z\sim6$ with constraints on the masses of their host galaxies, and will provide additional evidence for or against the evolution of the \magorrian relation with cosmic time.

For HSC J2239+0207 ($z = 6.25$) and two additional quasars, J073103.12+445949.4 ($z = 5.01$) and J134015.03+281328.1 ($z = 5.36$) (or J0731+4459 and J1340+2813; we use JHHMM+DDMM for brevity) we obtained NIRCam images and NIRSpec spectra. We search for host galaxy emission via PSF subtraction in the images, and in the case that the host galaxy is more compact than the core of the PSF, also set limits on the mass by searching for Balmer breaks in the integrated quasar-plus-host-galaxy spectra. The final two quasars,  SDSS J1148+5251 and ULAS J112001.48+064124.3 (J1120+0641) at $z\sim6.4$ and $7.1$ respectively, are among the highest-redshift quasars known. We obtained only NIRCam images of these quasars to search for hosts via PSF subtraction. 
For all these quasars, we determine host masses from our derived PSF-subtracted fluxes and compare them with values from the literature for galaxies without AGN.

We describe the process of reducing our NIRCam and NIRSpec data in Section \ref{sec:obs}. In Section \ref{sec:results}, we outline the steps taken to remove the quasar signal from the NIRCam images via PSF subtraction and place limits on the masses of the host galaxies. We also calculate black hole masses from the width of H$\beta$ and constrain Balmer break strengths for the quasars with NIRSpec observations. With these measurements, we discuss the location of our quasar sample relative to the local \magorrian relation in Section \ref{sec:discussion}, examine potential biases in this result, and outline the implications for the coevolution of SMBHs and galaxies at early times. We summarize our results in Section \ref{sec:summary}. Throughout this work, we assume a flat cosmology with $H_0 = 69.6$, $\Omega_M = 0.286$, and $\Omega_\Lambda = 0.714$. 

\section{Observations and Data Reduction} \label{sec:obs}

\subsection{NIRCam}

All our quasars were observed in three NIRCam filters chosen from F210M, F360M, F410M, F430M, and F480M (see Table \ref{tab:measurements}), with exposure times ranging from 265.3 to 2623.8 seconds depending on the quasar and filter. We used Module B SUB400P (FOV $12.5\arcsec \times 12.5\arcsec$, SW; $25\arcsec \times 25\arcsec$, LW) to image the quasar field using either the RAPID or BRIGHT2 readout modes. We adopted $4\times4$ dithering patterns to improve the PSF sampling and mitigate cosmic rays and detector artifacts, using primary dither type INTRAMODULEBOX and sub-pixel dither type STANDARD. Reference PSFs were obtained for all quasars by observing nearby bright stars. Stars were observed using the same module, dither types, and filter configurations as their corresponding quasars, but all stellar observations used the RAPID readout mode to avoid saturation. All stars were observed for 530.7 seconds in SW bands and 265.3 seconds in LW bands.

We processed our NIRCam data using the JWST pipeline version 1.9.6 \citep{Bushouse2023}, roughly following the procedures recommended in the STScI JWebbinars.\footnote{https://www.stsci.edu/jwst/science-execution/jwebbinars} The pipeline parameter reference file is registered in the JWST Calibration Reference Data System (CRDS) as jwst\_1084.pmap. We added a custom step to Stage 2 of the pipeline to characterize and subtract 1/f noise, a striping pattern in the images caused by the detector readout, from each frame. We produced final mosaic images of each quasar and star in Stage 3 of the pipeline by aligning and stacking all frames obtained in Stage 2, and resampled the images to smaller pixel scales (0.0147\arcsec/pixel in SW filters, 0.0300\arcsec/pixel in LW filters) using the drizzling algorithm in the Resample step of the pipeline to improve the accuracy of the PSF subtraction. Finally, we used the Photutils Background2D function on the mosaiced and resampled images to perform a global background subtraction.

\subsection{NIRSpec}

We also obtained NIRSpec/IFU data of three quasars in our sample, with a 3$\arcsec\times3\arcsec$ field of view composed of IFU elements of  $0.1\arcsec\times0.1\arcsec$ on the sky. J0731+4459 and J1340+2813 were observed with the G235M/F170LP disperser-filter combination,  offering wavelength coverage from 1.66 $\mu$m to 3.07 $\mu$m with a nominal spectral resolution of $\sim$1000; J2239+0207 was observed in the PRISM mode with a wavelength coverage from 0.60 $\mu$m to 5.30 $\mu$m with a nominal resolution $\sim$100. All these observations were carried out with the SPARE-CYCLING dither type in 
LARGE size at points 1, 2, 3 and 4 to mitigate cosmic ray and detector artifacts during data reduction. The NRSIRS2 readout was selected, with a total integration time
of 5894 seconds for J0731+4459, 7353 seconds for J1340+2813, and 8870 seconds for J2239+0207. 

We processed the NIRSpec IFU data with JWST pipeline version 1.12.0 \citep[][with CRDS pipeline parameter reference file jwst\_1088.pmap]{Bushouse2023} following the standard steps. This includes the first stage, \textit{Detector1Pipeline}, to apply detector-level corrections (e.g., dark current and bias subtraction, persistence correction, cosmic-ray removal) to the raw data of individual exposures and produce the uncalibrated 2D spectra; and \textit{Spec2Pipeline} to assign world coordinate system (WCS) information to the data, apply flat-field correction, make flux calibration, and construct 3D data cubes from the 2D spectra obtained at each dither location. The third stage, \textit{Spec3Pipeline}, combines the individual data cubes at different dither positions to produce the final merged data cube with outlier rejections, drizzling, etc. to remove any additional artifacts and improve spatial sampling. We found that the default pipeline did not always reject all obvious outliers: we therefore added an additional step to manually mask out the bad pixels in the 2D spectra after a careful visual inspection of individual cubes, and redid the final cube construction. Finally, we extracted the quasar spectrum using a circular aperture at a radius of 0.35\arcsec. For the background subtraction, we placed the same circular aperture at random locations of the cube that do not contain real structures, computed the median background spectrum, and subtracted it from the quasar spectrum.

\section{Analysis and Results} \label{sec:results}

\subsection{Point-spread function subtraction} \label{subsec:psfsub}

Our quasar observations have small fields of view with no bright stars, so building an empirical PSF from stars in the field \citep[as in e.g.,][]{Ding2023, Yue2023} was not a viable option. Instead, for each quasar we observed a nearby PSF star in the same bands immediately after observing the quasar (to reduce any temporal variation in the instrument's PSF between the images) and subtracted it from the quasar image to reveal any underlying host galaxy emission. 

Our approach has the advantage of being immune to different wavefront errors and PSF variation over the field of view. That is, the quasar and star are placed at the same location on the detector to minimize the effects from any spatial variation of the PSF across the detector. We additionally chose medium bands for all NIRCam observations to virtually eliminate differences in the profiles of the quasar and stellar PSFs due to their different spectral shapes across the filter (the flat continuum of the quasar versus the declining Rayleigh-Jeans tail of the star).  In a wide bandpass, the quasar's redder color in the filter could tend to broaden its PSF; if we then subtracted the narrower PSF of the star, we might observe a spurious galaxy detection.

However, we still considered the possibility of a discrepancy in the PSF shapes. We investigated the magnitude of any potential PSF mismatch by generating simulated stellar and quasar PSFs in each of our observed bands using WebbPSF. We used a Rayleigh-Jeans spectrum to generate the stellar PSFs, and a general quasar template \citep{Lyu2016}, shifted to the appropriate redshift, for the quasars. We constructed differential radial profiles of the stellar and quasar PSFs, and compute the normalized difference between the two as a function of radius. As in S23, we find no significant difference between the simulated quasar and stellar PSFs in any band. The normalized difference between the simulated stellar and quasar PSF at any given radius, $\frac{F(\mathrm{star})-F(\mathrm{QSO})}{F(\mathrm{star)}}$, is significantly less than $1\%$ in all bands and at almost all radii, except that normalized differences up to $5\%$ occur very near the PSF core where the annulus contains fewer pixels. We therefore do not expect any significant  residuals to be introduced in the PSF subtraction as a result of the different spectral shapes of the quasars and their PSF stars, particularly outside the PSF core in the region where we search for evidence of the host galaxy.

The process of PSF subtraction begins by aligning the background-subtracted star and quasar images in a given band. We perform this alignment by hand, iterating until minimal PSF artifacts are present in the subtracted images. We performed tests comparing the results of manual to computerized alignment: these indicated that the shifts obtained and the measured flux in the resulting PSF-subtracted images were identical whether the images were aligned by hand or automatically.

\begin{figure*}
    \centering
    \includegraphics[width=18cm]{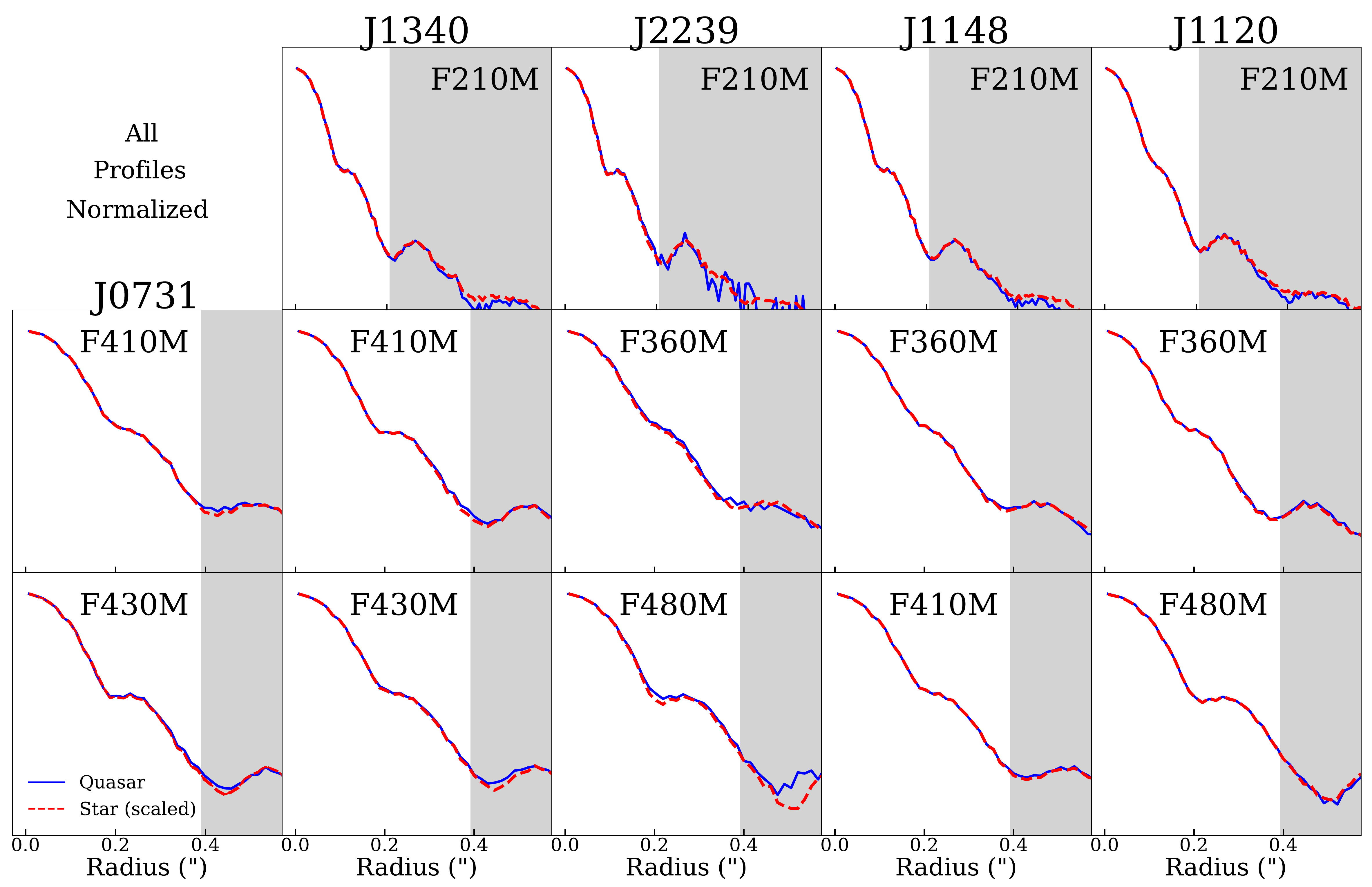}
    \caption{The normalized quasar and PSF star radial profiles. The quasars (columns) are shown in order of increasing redshift left to right, with the filter labeled. The quasar profile is shown as a blue solid line, and the radial profile of its corresponding PSF star, with the best-fit scaling to match the quasar profile at the center, is shown as a red dashed line. Each profile extends from the center of the image to a 0.6\arcsec radius: note that the F210M PSF is narrower than the other bands because of the smaller pixel scale. Most of the quasars and their PSF stars have virtually identical profiles out to $\sim13$ pixels from the center ($\sim0.4\arcsec$ in LW bands, bottom two rows, and $\sim0.2\arcsec$ in SW bands, top row), beyond which point noise drives any PSF mismatch (grey shaded region). The exception is J2239+0207 (center column) where small quasar-star PSF mismatches can be seen at $\sim0.2\arcsec$ in F360M and F480M. The excellent agreement between the quasar and stellar radial profiles underscores the efficacy of using medium-band filters for this analysis, and underscores the difficulty of detecting the host galaxy emission.}
    \label{fig:profiles}
\end{figure*}

Next, we normalize the stellar PSF image to match the flux of the quasar near the center of the image (within $\sim$ 0.15\arcsec), where the quasar light is most dominant. As we iterate the alignment and scaling, we examine the azimuthally-averaged radial profiles of the star and quasar, shown for all quasars in Figure \ref{fig:profiles}. If the star and quasar are properly aligned and scaled, their profiles will exhibit identical shapes at small radii. Indeed, as shown in Figure \ref{fig:profiles}, the profiles of the quasars and their PSF stars are virtually identical in all bands at small radii. Far from the center ($r\gtrsim0.4$\arcsec), deviations between the stellar and quasar PSFs are due to noise. The excellent agreement in the shape of the star and quasar radial profiles underscores the results from our WebbPSF simulations: despite their different spectral shapes, the quasars and their corresponding PSF stars exhibit nearly identical PSFs near the image centers. Any deviations between the PSF shapes of a quasar and star at slightly greater radii ($\sim$ 0.2--0.4\arcsec from the PSF center) are likely caused by extended emission from the host galaxy, rather than a PSF mismatch.

\begin{figure*}
    \centering
    \includegraphics[width=18cm]{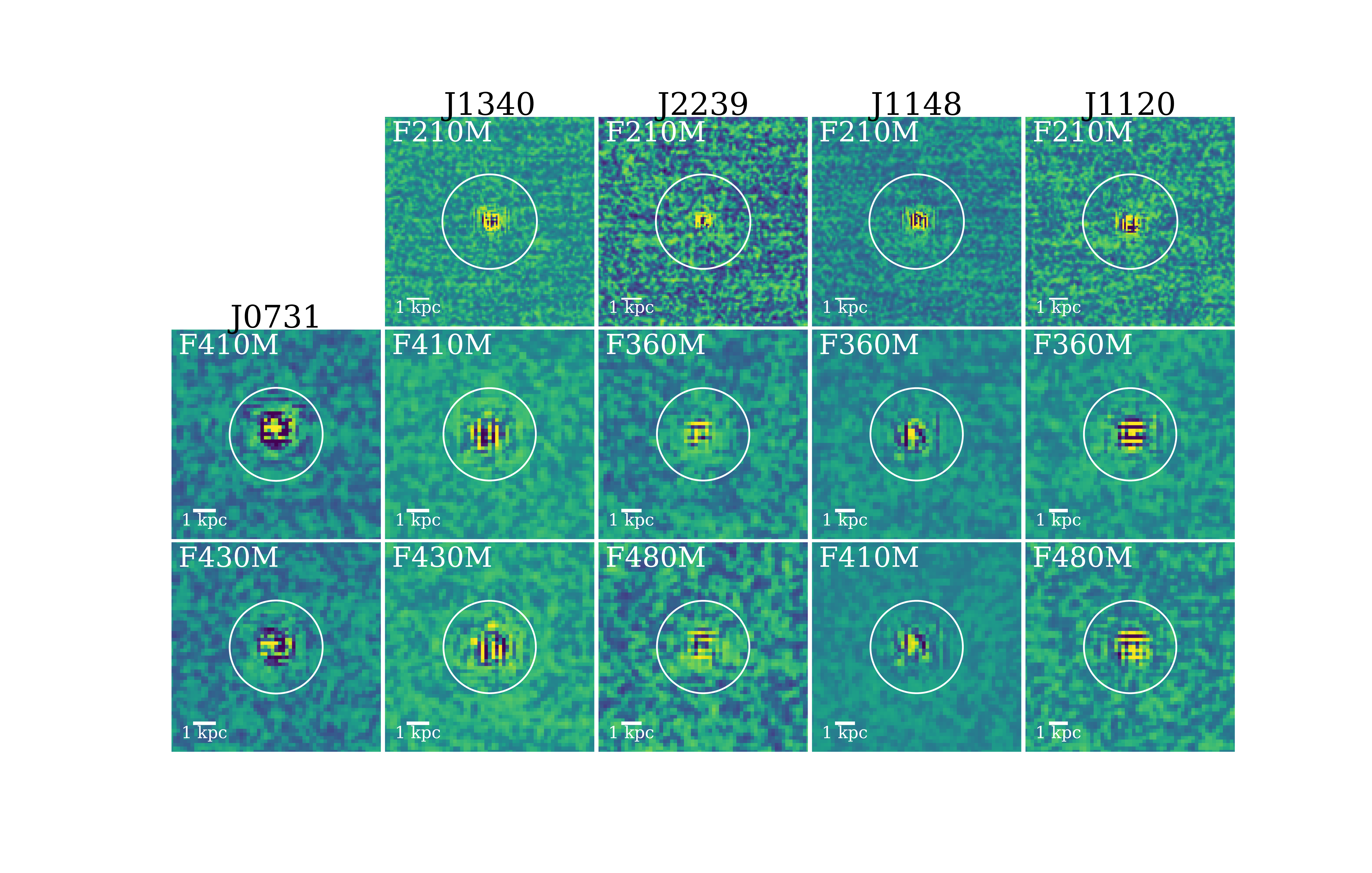}
    \caption{Our five observed quasars after subtraction of their stellar PSFs, normalized by the total (Poissonian + pixel) error budget. Each quasar (columns) is imaged in two or three bands between observed-frame 2.1 and 4.8 $\mu$m (rows). The 0.4\arcsec radius aperture used to extract galaxy fluxes and upper limits is shown as a white circle; each image has a field of view of $1.8$\arcsec. Consistent with the radial profiles (see Figure \ref{fig:profiles}), the only quasar that shows significant extended emission even when the reference PSF is normalized to the quasar flux at the center before subtraction is J2239+0207 (center column) in bands F360M and F480M.}
    \label{fig:psf-subtracted}
\end{figure*}

Once the stellar reference PSF is aligned and scaled to the quasar PSF, we subtract the stellar PSF and examine the residuals for evidence of extended emission. The PSF-subtracted images of all quasars are shown in Figure \ref{fig:psf-subtracted} in order of ascending redshift. The cleanliness of the images, with excess noise under the core of the PSF but few other residuals, demonstrates the quality of the subtractions. Only the F410M image of J0731+4459 displays a higher degree of PSF mismatch than the other images. Most of the images do not display evidence of extended emission, with the possible exceptions of J0731+4459 in F430M, J1340+2813 in F410M and F430M, and---as discussed in S23---J2239+0207 in F360M and F480M.

As a first estimate, we measure the PSF-subtracted flux by placing a $0.4\arcsec$ radius aperture on each PSF-subtracted image (we do not apply aperture corrections to these measurements, but based on the measured effective radii of typical $z\sim6$ galaxies from JWST \citep{Morishita2023} this aperture should capture $\gtrsim80\%$ of the galaxies' flux). The host galaxy of J0731+4459 ($z=5.01$, top row of Figure \ref{fig:psf-subtracted}) is undetected in the noisy F410M image (measured flux at the $<3\sigma$ level) and marginally detected ($\sim4\sigma$) in F430M. J1340+2813 ($z=5.36$, second row) is undetected in F210M and marginally detected (4-5$\sigma$) in F410M and F430M. J2239+0207 ($z=6.25$, third row) is undetected in F210M but returns fluxes at the $6.8$ and $5.5\sigma$ level in F360M and F480M respectively (as discussed in S23), our only apparently unambiguous detection with this PSF normalization method. J1148+5251 and J1120+0641 (bottom two rows, the two highest-redshift quasars in our sample) are undetected in all bands. That is, the only images in Figure \ref{fig:psf-subtracted} where the host galaxy emission can be unambiguously seen are in the F360M and F480M bands on J2239+0207. These results are consistent with the radial profiles (Figure \ref{fig:profiles}), where only J2239+0207 shows obvious deviations between the radial profiles of the quasar and its PSF star.

We report the flux in all images within a $0.4$\arcsec radius aperture in Column 5 of Table \ref{tab:measurements}.  This method of normalizing the reference PSF necessarily removes any flux from the host galaxy near the center of the image, and may lead to oversubtraction, potentially making the values underestimates, not true upper limits. Also, there are additional systematic errors associated with the PSF subtraction. We address these issues in the next section (\ref{subsec:sersics}), where we determine upper limits to the possible host galaxy fluxes to accompany the potentially oversubtracted fluxes derived here.

\subsection{Accounting for oversubtracted flux} \label{subsec:sersics}

\begin{figure*}
    \centering
    \includegraphics[width=15cm]{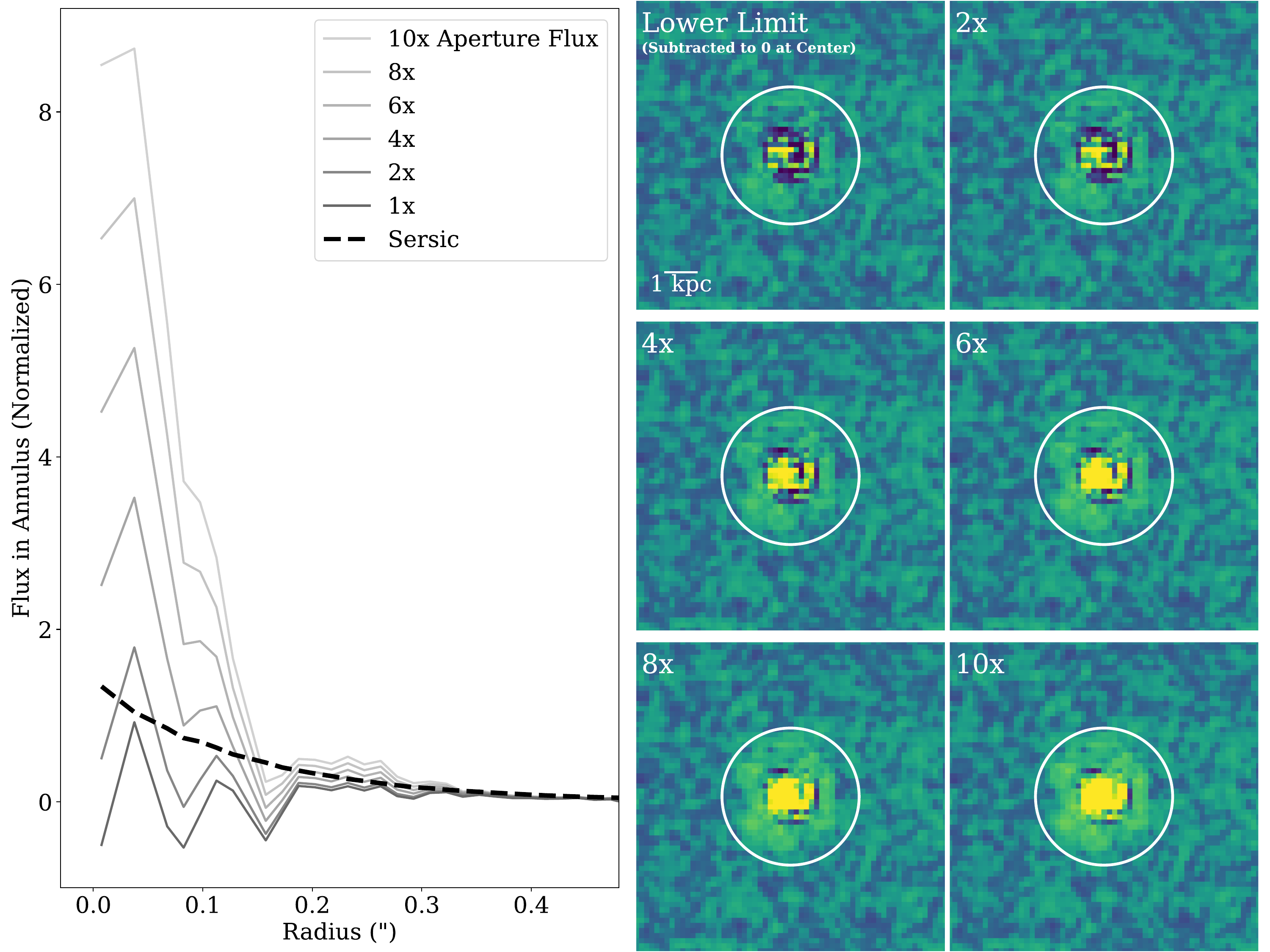}
    \caption{An illustration of the process of obtaining upper limits to the luminosities of our quasar host galaxies, with the test case of J1340+2813 in the F430M band (left). We subtract the images down to a range of flux levels, with the minimum being the flux obtained in a $0.4$\arcsec aperture when the reference (stellar) PSF is normalized to the quasar PSF (as in Figure \ref{fig:profiles}). At right, we show the resulting ``PSF-subtracted" images: clearly, remnants of the PSF pattern begin to become visible when the flux in the aperture is tuned to $4-6$ times the normalized-to-zero flux. At left, we show the corresponding radial profiles of the PSF-subtracted images. At $>6x$ the lower limit flux, the radial profile is no longer consistent with an $n=1$ S\'ersic profile (black dotted line). We take the maximum flux where the PSF pattern is not obvious in the images and the radial profile is consistent, on average, with the S\'ersic at $r>0.1$\arcsec as the upper limit on the galaxy flux (column 6 of Table \ref{tab:measurements}).}
    \label{fig:uplims}
\end{figure*}

Given how faint the host galaxies are, the dominant errors in measuring them (or obtaining limits) are the systematic ones associated with the PSF subtraction, not statistical errors. In addition, because the host galaxies of our quasars are very faint, the usual practice of subtracting the PSF down to the projected galaxy emission is not feasible. Instead, as described in the preceding section, we scaled the stellar PSF to match the quasar PSF at the center, which necessarily reduces the residual flux to zero near the center of the image. Placing an aperture on the PSF-subtracted image therefore returns essentially a lower limit on the total galaxy flux, because any flux at the very center of the galaxy (behind the PSF artifacts) has been subtracted. To estimate the amount of ``missing" flux and provide an alternate, more conservative measurement of the galaxy flux and therefore its mass, we iterate the normalization of our reference PSF and fit S\'ersic profiles to the residual flux in the PSF-subtracted images.

We begin by modifying the normalization of the reference PSF. When we reduce the normalization of the reference PSF to the quasar PSF, the resulting profile is broadly consistent with a S\'ersic profile for all of our quasars, especially at $r>0.2$\arcsec (see Figure \ref{fig:uplims}). Early JWST results \citep[e.g.][]{Yang2022, Ormerod2023} have found that S\'ersic indices of $n\sim1$ are typical for galaxies at $z\sim6$: indeed, we find that this is the S\'ersic index most consistent with the profiles of our PSF-subtracted images. While a higher S\'ersic index may provide a slightly better fit to the very inner, extremely noisy part of the image (at $r<0.1$\arcsec), any index above $n\sim1.5$ does not fit the region from $0.15$\arcsec to $0.4$\arcsec well. The quasar PSF, on the other hand, is highly inconsistent with a S\'ersic profile. Therefore, by decreasing the flux of the reference PSF before subtracting and examining the resulting modified PSF-subtracted image and radial profiles until the PSF pattern becomes visible in the ``subtracted" image and the radial profile is no longer well-fit by an $n=1$ S\'ersic profile, we can place an upper limit on the flux attributable to the galaxy rather than to the PSF. We adjust the normalization to a range of values, such that the measured flux in the $0.4$\arcsec aperture returns from two to twenty times the flux obtained when normalizing the stellar PSF to the quasar PSF at the center of the image. An example of this process is shown in Figure \ref{fig:uplims}, for the F360M image of J1340+2813.

We examine the radial profiles of these adjusted PSF-subtracted images, and fit the region of the profile relatively unaffected by central PSF artifacts (at radii $\gtrsim0.15$\arcsec), with an $n=1$ S\'ersic profile. We set the effective radius of the model S\'ersic using equations 5 and 6 of \cite{Morishita2023}, determining an ``expected" mass from the local \magorrian relation and the black hole masses of our quasar sample: these effective radii are between 1 and 1.5 kpc for our quasars.

We find only small adjustments to the normalization are needed to account for oversubtraction. For most bands, when tuned to return more than $4-6$ times the original, normalized-to-zero-at-center flux, the PSF pattern begins to become apparent in the subtracted images, and the core of its radial profile becomes inconsistent with a S\'ersic profile (see Figure \ref{fig:uplims}). The exception is the F410M image of J0731+4459, which displays worse stellar-quasar PSF agreement than the rest of the sample and therefore a noisier radial profile; we can increase its flux by approximately a factor of $20$ before the radial profile becomes inconsistent with the S\'ersic. We report the upper limit on the J0731+4459 F410M flux in Table \ref{tab:measurements}, but do not use this band to constrain the galaxy mass. We take the maximum normalization that returns a) a subtracted image without obvious residual PSF patterns, and b) a radial profile that can be fit by an $n=1$ S\'ersic profile, as the upper limit on the flux of the galaxy as listed in column 6 of Table \ref{tab:measurements}.

\begin{figure*}
    \centering
    \includegraphics[width=18cm]{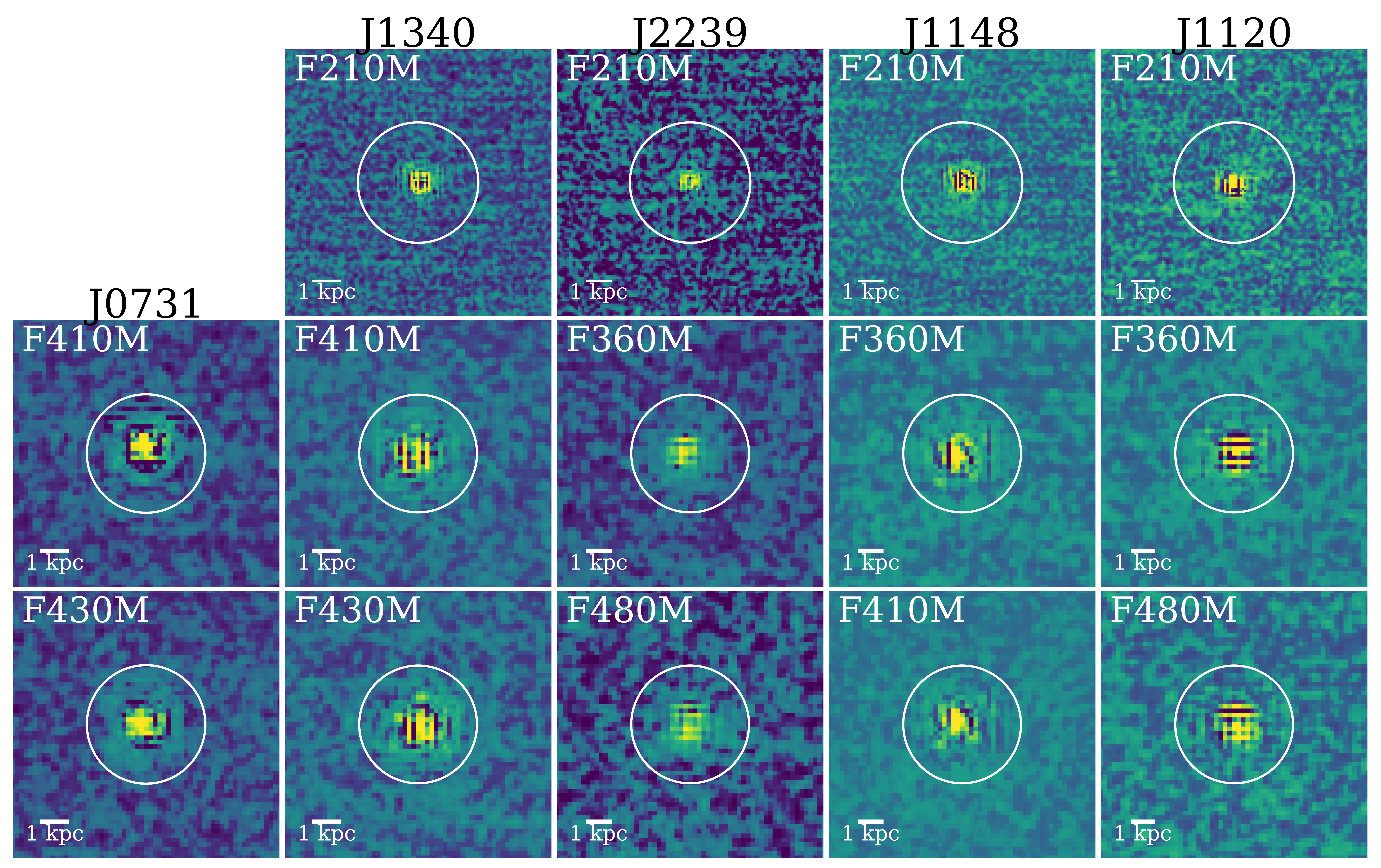}
    \caption{Same as Figure \ref{fig:psf-subtracted}, but with the normalization of the reference PSF to the quasar image adjusted as described in Section \ref{subsec:sersics} in order to calculate the upper limits on the galaxy flux. The 0.4\arcsec radius aperture used to extract galaxy fluxes is shown as a white circle; each image has a field of view of $\sim1.8$\arcsec. For comparison, each image is shown with the same normalization and stretch as in its corresponding panel in Figure \ref{fig:psf-subtracted}.}
    \label{fig:psf-subtracted_alt}
\end{figure*}

\subsection{Galaxy Masses} 

Our five quasars possess high-mass supermassive black holes ($\log (M_{\mathrm{BH}} \, [M_{\odot}]) = 8.8$ to $9.8$): the local \magorrian relation therefore predicts large host galaxy masses, between $\log(M_* \,[M_{\odot}])=11$ and $12$. Such massive hosts would have been easily detected. However, obtaining realistic host galaxy masses is nontrivial, as each quasar has been observed in no more than three closely spaced bands. It is therefore not feasible to determine the host masses by performing e.g. spectral energy distribution (SED) fitting because of the degeneracies in models due to our small number of data points, combined with the low signal to noise. We must therefore turn to another method.

The Spitzer Space Telescope observed hundreds of high-redshift AGN and inactive galaxies during its mission, including many galaxies in legacy fields with additional data spanning the UV to far-IR or radio, and therefore with robust masses from SED fitting. We can therefore use these galaxies to examine the relationship between their IRAC Band 1 (3.6 $\mu$m) fluxes, whose bandpass closely resembles NIRCam F360M, and their stellar masses from SED fitting. At the redshifts of our quasars, F360M probes the rest-frame optical and should therefore scale with the galaxy mass: we can use this to build an empirical relationship between 3.6 $\mu$m flux and mass to constrain the masses of our quasar host galaxies. This method is based on the assumption underlying the Magorrian relation that quasar host galaxies, as a population, should resemble field galaxies at the same redshift.

\cite{Mclure2011}, \cite{Curtis2013}, and \cite{Jiang2016} model massive $z\sim5-8$ galaxies with high-quality observations spanning a wide range of wavelengths, including IRAC Band 1 measurements.\footnote{We use the exponentially decaying star formation models of \cite{Mclure2011}, models E of \cite{Curtis2013}, and the \cite{Jiang2016} models including nebular emission.} These samples include high-mass, highly star-forming galaxies similar to those we expect to host our luminous quasars. Because these sample galaxies---as well as our quasar hosts---lie at a range of redshifts, the IRAC Band 1 filter probes a range of rest-frame wavelengths. This may introduce biases if the galaxies' spectra are not flat, requiring complicated K-corrections. Reassuringly, however, when we create model SEDs with reasonable star formation histories and the small levels of reddening typical of galaxies at these redshifts (E(B-V) $\sim$ 0.05), we find that they are virtually flat in frequency units at the wavelengths of interest (rest-frame $\sim$ 450-600 nm). The necessary K-corrections are therefore negligible.

\begin{figure*}
    \centering
    \includegraphics[width=12cm]{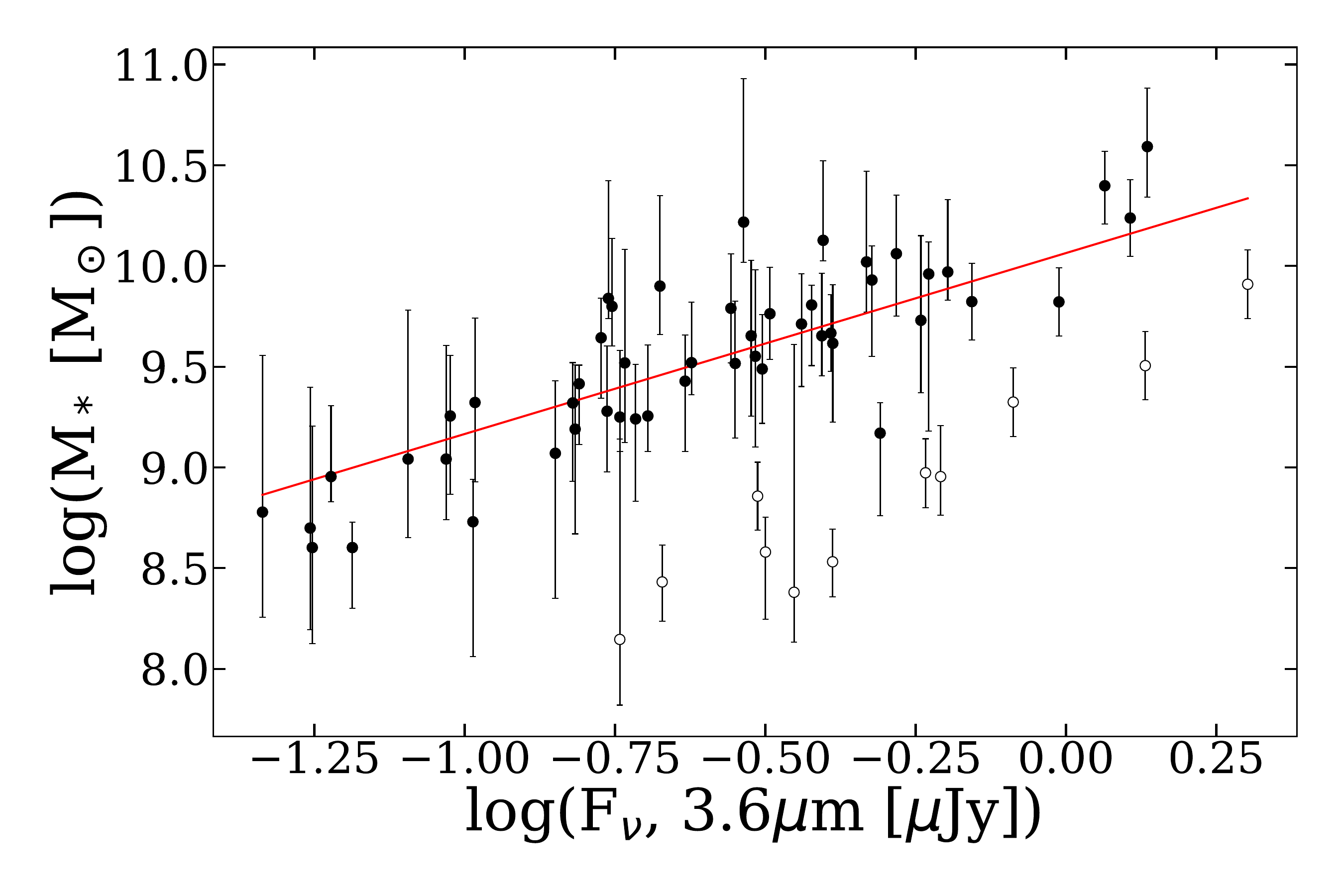}
    \caption{Galaxy masses vs. IRAC Band 1 flux density converted to equivalent values at z = 6, and the best-fit model (red line, Equation \ref{eq:masses}). Models from \citet{Jiang2016} that derive ages of the dominant stellar population $<$ 10 Myr, which we exclude from the fit, are shown as open circles. }
    \label{fig:masses}
\end{figure*}

We correct the IRAC Band 1 fluxes of the \cite{Mclure2011}, \cite{Curtis2013}, and \cite{Jiang2016} to a single redshift ($z=6$), and plot their masses as a function of IRAC Band 1 flux in Figure \ref{fig:masses}. The errors on the \cite{Mclure2011} and \cite{Curtis2013} galaxies are directly from the references, but the quoted errors in \cite{Jiang2016} are much smaller and appear not to include systematic uncertainties. We therefore increase the error bars on the \cite{Jiang2016} points by 0.15 dex, making them similar to the errors quoted in the other references to ensure those points do not dominate our fits. 

The points in Figure \ref{fig:masses} clearly fall into two distinct populations. Very young galaxies, with a dominant stellar population age (from SED fitting) less than 10 Myr have a larger 3.6 $\mu$m flux for a given mass: these are plotted as open circles in Figure \ref{fig:masses}. Older galaxies, on the other hand, which we expect to more closely resemble the galaxies hosting our quasar sample, produce less 3.6 $\mu$m flux at a given mass (filled circles in Figure \ref{fig:masses}). We want to place maximally conservative upper limits on the masses of our host galaxies; including the very young galaxies biases our fit, tending to return a lower galaxy mass for a given flux. Using such a fit may lead us to mistakenly conclude that our quasar host galaxies lie above the local \magorrian relation even if they do not. A more conservative result is obtained by rejecting these very young galaxies when we fit the relationship between stellar mass and 3.6 $\mu$m flux.

We therefore fit the relationship between the $3.6 \mu$m flux density (adjusted to $z=6$) and galaxy mass measurements by least squares minimization, holding the slope to 1 (i.e., the mass should be proportional to the $3.6\mu$m flux density) and excluding the galaxies with dominant stellar population age $<10$ Myr. The reduced $\chi^2$ for this fit is 1.98; if the single most discordant point (the value at $\log(F_{\nu} \; [\mu \mathrm{Jy}]) = -0.31$, $\log(M_* \, [M_{\odot}]) = 9.17$) is rejected, the reduced $\chi^2$ is 1.50. The equation of this best-fit line is

\begin{equation}
\log(M_* \, [M_{\odot}]) = \log(F_{\nu} \; [\mu \mathrm{Jy}]) + 10.20,
\label{eq:masses}
\end{equation}

\noindent
where $F_\nu$ is the 3.6 $\mu$m flux adjusted to $z=6$ in $\mu$Jy and the galaxy mass is in solar units. 

With this empirical relation, we can place lower and upper limits on the masses of our quasar host galaxies from the lower and upper limits on their fluxes. As with the galaxies in the models, we begin by correcting our lower limits (from aperture photometry, column 5 of Table \ref{tab:measurements}) and upper limits (from varying the normalization in Section \ref{subsec:sersics}, column 6 of Table \ref{tab:measurements}) on the galaxy fluxes to $z=6$; we again make the assumption that the galaxy SED is flat at the wavelengths of interest, around rest-frame 500 nm. We can then use these K-corrected fluxes and Equation \ref{eq:masses} to obtain a mass range for each quasar host galaxy. For the galaxies with an F360M flux (J2239+0207, J1148+5251, and J1120+0641), we utilize that flux to calculate the mass. For J1340+2813, we use the band at the most similar wavelengths, F410M. J0731+4459 also has an F410M flux, but because its F410M PSF-subtracted image displays more severe PSF subtraction artifacts than its F430M image, suggesting some degree of mismatch between its quasar and stellar PSF in that band, we conservatively use F430M to calculate its mass instead. Again, the rest-frame optical SED is very flat for these galaxies and using F410M or F430M instead of F360M should not probe a significantly different part of the SED. The masses are reported in Column 7 of Table \ref{tab:measurements}.

\subsection{BH masses and Balmer breaks from NIRSpec spectra}

\begin{figure*}
    \centering
    \includegraphics[width=16cm]{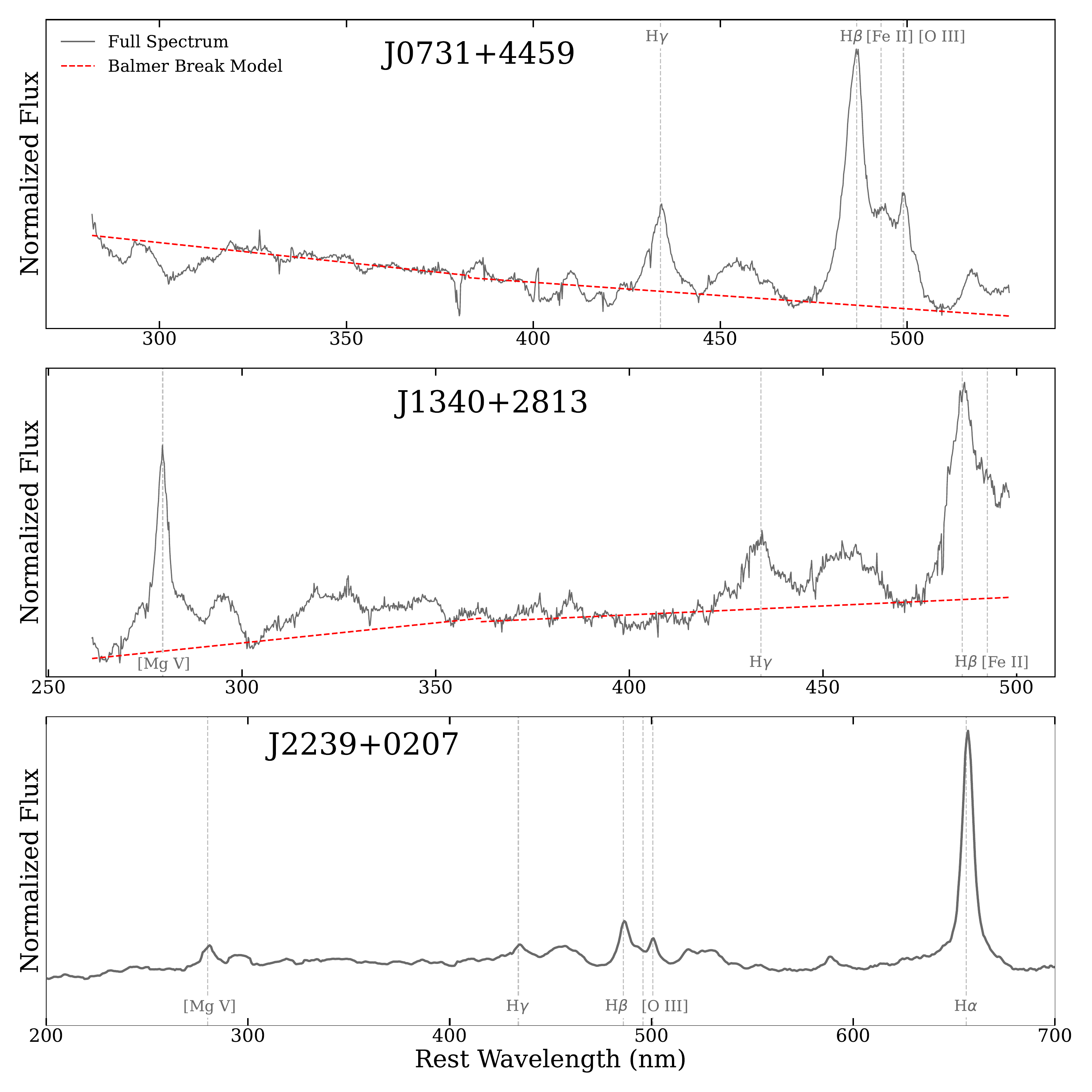}
    \caption{The NIRSpec spectra of three of our quasars: J0731+4459 (top row, $z=5.01$), J1340+2813 (center, $z=5.36$), and J2239+0207 (bottom row, $z=6.25$). Prominent spectral lines are labeled. We fit the spectra of J0731+4459 and J1340+2813 on either side of the Balmer break to constrain its strength, masking out emission and absorption lines, and show the best-fit model as a red dashed line. The scale for these two quasars is amplified and the zero point suppressed to emphasize the small size of the resulting fitted break. Neither galaxy displays an obvious Balmer break; the spectra of all three sources are very typical of high-redshift quasars.}
    \label{fig:spectra}
\end{figure*}

We obtained NIRSpec spectra of three of our quasars: J0731+4459 ($z=5.01$), J1340+2813 ($z=5.36$), and J2239+0207 ($z\sim6.25$) to constrain the strength of the Balmer break and measure black hole masses from H$\alpha$ and H$\beta$. All three quasar spectra are shown in Figure \ref{fig:spectra} in order of increasing redshift. The prism spectrum of J2239+0207 (bottom row of Figure \ref{fig:spectra}) is extensively discussed in S23: the spectral resolution and signal-to-noise ratio are low, but we are nonetheless able to use H$\alpha$ to place a lower limit on the black hole mass of $\log (M_{\mathrm{BH}} \, [M_{\odot}]) > 8.5$, consistent with previous measurements from Mg~II and C~IV. The redshifts of both J0731+4459 and J1340+2813 place H$\beta$ within the NIRSpec wavelength range, which we use to calculate an additional black hole mass estimate.

We do not observe an obvious narrow component of H$\beta$ in the spectra of J0731+4459 and J1340+2813. However, both spectra display a [Fe~II] emission feature blended with H$\beta$, which must be fit simultaneously with the H$\beta$ line. We therefore fit two Gaussian profiles---for the broad component of H$\beta$ and for the [Fe~II] feature---superimposed on a local linear continuum. We fix the central wavelengths to the line centers and allow the amplitudes and widths of both lines to vary. For J0731+4459, we obtain a FWHM of $\sim3400$ km/s. This line width can be converted to a black hole mass using Equation 5 in \cite{Vestergaard2006}:

\begin{multline} \label{eq:bhmass}
    \log{M_{\mathrm{BH}}} = \log \Biggl \{ \! \biggl [ \frac{\mathrm{FWHM(H}\beta)}{1000 \; \mathrm{km \, s}^{-1}} \biggr ]^{2}   \biggl [ \frac{\lambda L_{5100 \mbox{\scriptsize\normalfont\AA}}}{10^{44} \; \mathrm{erg \, s}^{-1}} \biggr ]^{0.5} \Biggr \} \\ + (6.91 \pm 0.02)
\end{multline}

\noindent
where we use the flux density in F410M (Table \ref{tab:measurements}) as a proxy to determine the continuum luminosity at $5100\AA$. We find a black hole mass $\log (M_{\mathrm{BH}} \, [M_{\odot}]) = 9.3^{+0.2}_{-0.3}$.

For J1340+2813, we measure a FWHM of 5100 km/s, and using Equation \ref{eq:bhmass} obtain a mass $\log (M_{\mathrm{BH}} \, [M_{\odot}]) = 9.4^{+0.2}_{-0.4}$. This value is consistent with that reported by  \citet{Jun2015}.

J0731+4459 and J1340+2813 also have rest-frame $3650\angstrom$ positioned within the NIRSpec medium wavelength range. While their host galaxies are only marginally detected in the NIRCam images, their spectra might reveal a Balmer break if there is a massive host galaxy sufficiently compact to hide behind the PSF subtraction artifacts.

To test for this possibility, we fit a Balmer break spectrum to  the quasar spectra, varying the strength of the break. We created a model of the Balmer break using the
PopStar models \citep{Molla2009} assuming a Chabrier initial mass function. The strength of any Balmer break depends on the age of the stellar population dominating the SED; a very young population will have a very weak feature. For a baseline model, we have assumed constant star formation for 500 Myr. To fit the resulting complex spectrum to the quasar spectrum, we first mask all absorption and emission features in both spectra and fit the model PopStar spectrum with two lines, shortward and longward of the Balmer break, and a cubic function connecting the two to represent the break itself. We fit this  parameterization of the Balmer break to the spectra of our quasars, with the normalization of the model, the magnitude of the break, and an overall slope (to represent any effects of dust extinction) as free parameters. We allowed negative break values as a test for fitting residuals.

The best-fit strengths of the Balmer break are approximately $0.8\%$ and $1.4\%$ the flux of the quasar continuum in J0731+4459 and J1340+2813, respectively, as shown in Figure~\ref{fig:spectra}, and both are negative (i.e. the fit returns a {\em reverse} break, with the spectrum brighter at blue wavelengths than red). We do not believe this is significant, as in both cases, no break at all is consistent with the spectra and fitting a single flat continuum returns a nearly identical reduced $\chi^2$.

Placing an upper limit on the Balmer break is therefore difficult, because the best-fit value of the break is negative and consistent with zero: clearly the contribution of the stellar continuum to the spectrum is very small. As a rough estimate of the maximum possible contribution, we take upper limits to the nominal break values as the negatives of the best fits (so the breaks are positive). These stellar continuum fluxes permit galaxy flux densities in J0731+4459 and J1340+2813 of 2.5 and 5.4 $\mu$Jy respectively in the 3 to 4 $\mu$m bands, values consistent with the upper limits in Table \ref{tab:measurements} for the same quasars. This result, and the lack of obvious stellar contribution when visually inspecting the quasar spectrum, implies that there is not significant flux (relative to the quasar flux) from the galaxy in the central regions of the PSF, and rules out the possibility of  compact host galaxies sufficiently massive to preserve the local \magorrian relation hiding behind PSF subtraction residuals in the core of the PSF.

\begin{deluxetable*}{ccccccc}
\tablecaption{Measured properties of the high-$z$ quasar sample \label{tab:measurements}}
\tablewidth{0pt}
\tablehead{
\colhead{Quasar} & \colhead{Redshift} & \colhead{Band} & \colhead{Quasar} & \colhead{Measured} & \colhead{Upper Limit} & \colhead{Inferred} \\[-0.25cm]
\colhead{} & \colhead{} & \colhead{} & \colhead{Flux} & \colhead{Galaxy Flux} & \colhead{Galaxy Flux} & \colhead{Galaxy Mass} \\
\colhead{} & \colhead{} & \colhead{} & \colhead{($\mu$Jy)} & \colhead{($\mu$Jy)} & \colhead{($\mu$Jy)} & \colhead{$\log$(M$_{\odot}$)}
}
\decimalcolnumbers
\startdata
J0731+4459 & 5.01 & F410M & $194.0\pm0.2$ & $0.24\pm0.21$ & $4.32$ & \\
 &  & F430M & $112.0\pm0.2$ & $0.85\pm0.22$ & $5.12$ & $<10.7$\\
\hline
 &  & F210M & $95.1\pm0.3$ & $0.78\pm0.28$ & $1.56$ & \\
J1340+2813 & 5.36 & F410M & $173.2\pm0.2$ & $0.88\pm0.18$ & $5.28$ & $<10.8$ \\
 &  & F430M & $192.7\pm0.2$ & $1.06\pm0.23$ & $6.36$ & \\
\hline
 &  & F210M & $6.81\pm0.04$ & $0.12\pm0.07$ & $0.60$ & \\
J2239+0207 & 6.25 & F360M & $7.43\pm0.05$ & $0.34\pm0.05$ & $1.36$ & $10.0^{+0.4}_{-0.3}$ \\
 &  & F480M & $11.32\pm0.09$ & $0.45\pm0.08$ & $1.32$ & \\
\hline
 &  & F210M & $120.5\pm0.2$ & $0.39\pm0.23$ & $0.78$ & \\
J1148+5251 & 6.40 & F360M & $117.1\pm0.2$ & $0.45\pm0.18$ & $2.70$ & $<10.7$ \\
 &  & F410M & $103.2\pm0.1$ & $0.48\pm0.11$ & $1.92$ & \\
\hline
 &  & F210M & $42.0\pm0.1$ & $0.12\pm0.12$ & $0.48$ & \\
J1120+0641 & 7.09 & F360M & $53.2\pm0.1$ & $0.20\pm0.07$ & $0.80$ & $<10.3$ \\
 &  & F480M & $48.3\pm0.1$ & $0.31\pm0.13$ & $0.62$ & \\
\enddata
\tablecomments{The measured quasar and host galaxy fluxes of our five high-$z$ quasars. The measured galaxy fluxes (5) were obtained by normalizing our reference PSF to match the flux of the quasar at the center of the image before subtracting, therefore subtracting the flux to zero near the center of the host galaxy (see Section \ref{subsec:psfsub}). The uncertainties in (5) reflect purely statistical errors, without taking into account systematics.} We then adjusted the normalization of the reference PSF to determine how much flux could be attributed to the host galaxy near the image center, and adopted the maximum flux with subtracted image and radial profile consistent with a S\'ersic profile rather than a PSF as the upper limit on the galaxy flux (6). We then calculate masses (7) by applying Equation \ref{eq:masses} to the limits on the host galaxy flux.
\end{deluxetable*}

\section{Discussion} \label{sec:discussion}

The existence of the supermassive black holes in very high redshift, very luminous quasars is a severe test of conventional theories that supermassive black holes and their host galaxies grow symbiotically from very early times. This issue is most challenging for the most massive SMBHs, e.g. those in luminous quasars. Our quasar host galaxy masses, reported above, appear to challenge this hypothesis, and imply that the local \magorrian relation is not well established at $z\sim5-7$. This section explores this possible contradiction in more depth. We begin by discussing theoretical predictions of host galaxy masses for quasars similar to our sample and comparing these predictions to our measurements. We then discuss whether our results could be missing the host galaxies or underestimating their masses because of extinction by interstellar dust. We conclude by showing that the deviation of our and other $z\sim5-7$ quasar samples from the local \magorrian relation is unlikely to be a consequence of selection effects.

\subsection{Testing Quasar-Galaxy Symbiotic Evolution}

A theoretical example of the growth of a galaxy and SMBH together is provided by \cite{Li2008} (see their Figure 1). In these particular models, the central supermassive black hole and its host galaxy have reached their final masses by $z=5$, with nearly all growth (particularly of the host galaxy) complete even earlier, by $z=6$. As such, between $z=6$ and $z=5$, the host galaxy is slightly overmassive compared to the black hole. 

\cite{Habouzit2022} discuss the results of six large cosmological simulations with different approaches to SMBH-galaxy coevolution, providing a more complete view.  Such studies are limited by the currently achievable simulation volumes (see \citet{Habouzit2022} for a discussion), but the two with  adequate statistics to extend to high-luminosity quasars indicate that at $z=5$ they reside in host galaxies in the $0.3$ to $1 \times 10^{11}$ M$_\odot$ range, with relatively little scatter\footnote{For the TNG300 case, the probability density is $>$ 0.1 for 10.27 $<$ log($M_*/M_\odot$) $<$ 10.77 and for SIMBA it is $>$ 0.1 for 10.68 $<$ log($M_*/M_\odot$) $<$ 11.10.}.
However, the quasars we have studied have SMBH masses  well above the simulation results\footnote{The {\it maximum} SMBH masses in the simulations are $< 6 \times 10^8$ M$_\odot$.}. J2239+0207 lies at the very high end of the simulated distribution (S23), but our other four quasars have much larger SMBH masses: for J0731+4459, $2 \times 10^9$ M$_\odot$ (this paper); J1120+0641, $1.35 \times 10^9$ M$_\odot$ \citep{Yang2021}; J1148+5251, $7 \times 10^9$ M$_\odot$, \citep{Barth2003, Willott2003,Shen2019}\footnote{There is a wide range of estimates for this quasar's black hole mass, with \cite{Willott2003} estimating $\sim 3 \times 10^9$ M$_\odot$ and \cite{Shen2019} $9.9 \times 10^9$ M$_\odot$ from Mg II and $10.2 \times 10^9$ M$_\odot$ from C~IV. There is no evidence of variability in the J-band magnitudes from \cite{Willott2003} and \cite{Shen2019} that could explain this difference. The calibration of line width SMBH masses has changed modestly over time \citep{Jun2015}, which could account for a small part of the difference. In averaging the values we have therefore given slightly greater weight to the \citet{Shen2019} ones.}; J1340+2813, $2.5 \times 10^9$ M$_\odot$, \citep[][this paper]{Jun2015}. The additional quasars analyzed for host galaxies by \cite{Yue2023} also all have very massive SMBHs, as listed in that paper, ranging from $1.9\times10^9$ to $1.2\times10^{10}$ M$_\odot$. The AGN luminosities are proportional to their SMBH masses \citep{Shen2019}, placing all of these quasars high into the ``bright'' category \citep{Habouzit2022}. According to the  predictions from existing simulations, we therefore expect these SMBHs to lie in hosts of $1 \times 10^{11}$ M$_\odot$ (and larger), significantly more massive than the upper limits we have calculated for our quasar sample. 
We now consider whether the apparent contradiction to these predictions we have found might arise from other effects than the intrinsic quasar/host behavior.

 \subsubsection{Effect of Intense Nuclear Star Formation}

Three members of our sample, J1148+5251, J1340+2813, and J2239+0207, have very large far-infrared luminosities indicative of star formation rates $>>$ 1000 M$_\odot$ yr$^{-1}$ \citep{Leipski2014, Izumi2019, Bertoldi2003}. Such star forming regions are expected to be centrally concentrated within the host galaxy \citep[e.g.,][]{Lapi2018} and their frequent occurrence around high-redshift quasars may be directly related to the growth of the SMBHs and the energy output from the quasar \citep[e.g.,][]{Hopkins2010, Li2022}. The regions are sufficiently luminous and compact to approach the Eddington limit, where the radiation pressure from the young, hot stars blows the dust and with it the interstellar gas out of the region and quenches the star formation \citep{Thompson2005,Crocker2018}. As a result, these extremely high levels of star formation observed in J1148+5251, J1340+2813, and J2239+0207 are likely very short-lived; nonetheless, they can contribute significantly to the mass of the host galaxy (e.g., 5 Myr of creation of 2000 M$_\odot$/yr$^{-1}$ of stars yields $10^{10}$ M$_\odot$ of stars for a conventional, i.e. Chabrier or Kroupa, initial mass function, although the actual mass function in these regions is probably more top-heavy and the mass yield therefore lower). Repetitive short bursts of this nature quenched by the Eddington limit process are proposed to be a mechanism for feeding black holes and activating AGN \citep{Davies2007}.

Although important to understand the co-evolution of the quasars and their host galaxies, the gas, dust, and star formation described above is centrally concentrated within a radius $\lesssim$ 1 kpc \citep{Lapi2018}, in the region made invisible in our images by the quasar and the PSF subtraction artifacts. Our galaxy masses and upper limits are derived for the extended, older, more quiescent part of the galaxy outside this central region. 

\subsubsection{Host Galaxy Extinction}

Our failure to detect host galaxies might be attributed to galaxy-scale obscuration by interstellar dust. Our quasars fall into two broad categories. In J0731+4459 and J1120+0641 there is no evidence for ongoing strong nuclear star formation and the accompanying gas and dust and the extinction should be typical of $z\sim6$ field galaxies. As already discussed, J1340+2813, J2239+0207, and J1148+5251, however, have very high far infrared luminosities and should have significant obscuration in their central regions. We begin by considering the first category. 

\vspace{3mm}

\noindent
{\it Extinction in typical $z\sim6$ galaxies:} The galaxies in Figure~\ref{fig:masses}, used to derive Equation \ref{eq:masses} and determine host galaxy masses, in general show very low levels of extinction. Is this a reasonable assumption to make for typical $z\sim6$ galaxies in general, and the subset of our quasar sample without high levels of nuclear star formation? The galaxies from \cite{Mclure2011}, \cite{Curtis2013}, and \cite{Jiang2016}, predating JWST, were by necessity selected on the basis of the visible and near infrared measurements and therefore may be biased toward low extinctions. To first order, this selection does not undermine their use as mass indicators, but it could affect the application of Equation~\ref{eq:masses} to other galaxies at $z\sim6$, since they might have systematically higher extinction.  

\cite{Nakajima2023}, using a mix of direct-temperature and strong-line metallicity calibrations, find typical metallicities of 12 + log(O/H) $\sim$ 8.2 for galaxies at $z\sim6$ with masses of $\sim 10^{10}$ M$_\odot$ (i.e., similar to the quasar hosts); this is a metallicity four times lower than the solar value \citep[8.8,][]{vonSteiger2016}. The ratio of dust to gas mass is then expected to be an order of magnitude lower than in galaxies of approximately solar metallicity \citep{Remy2015,Aniano2020}. It is therefore not surprising that the extinctions are low in galaxies at $z\sim6$, and the sources in Figure \ref{fig:masses} are likely not significantly biased towards uncommonly low extinction. 

The levels of extinction for infrared-selected  galaxies (i.e., observed with JWST NIRCam) are discussed in \citet{Gomez2023} for a sample at $3 < z < 7.5$. They find that about 76\% of their sample has $A_V < 0.5$ (probed by the F150W$-$F444W color) and another 17\% has $0.5 < A_V < 1$. These extinction levels are only slightly higher than those for the sample shown in Figure~\ref{fig:masses}. Assuming that the two quasars in our sample without large far-infrared excesses lie in relatively typical $z\sim6$ galaxies, these low extinction levels support our assumption that the galaxy-scale interstellar extinction is low in J0731+4459 and J1120+0641.

\vspace{3mm}

\noindent
{\it Extinction in highly star-forming $z\sim6$ galaxies:} However, the three other quasars in our sample (J1340+2813, J2239+0207, and J1148+5251) with huge far-infrared luminosities may be highly-dusty exceptions to the rule of generally low interstellar dust and extinction at $z\sim6$. We examine this possibility using a sample of local LIRGs and ULIRGs, which, like our three quasars, exhibit large far-infrared excesses compared to typical galaxies at their redshift \citep[by a factor of a few to tens: ][]{Howell2010}. We use the $J-K$ color difference to judge the extinction: these longer-wavelength bands trace old stars and this color is therefore relatively immune to differences in star formation history. We draw examples from the multi-aperture $JHK$ photometry in \citet{Smith1996}, selecting galaxies with infrared luminosity $> 3 \times 10^{11} L_{\odot}$ \citep{Sanders2003}. We calculate the color difference in all cases for annuli ranging from 5 to $20\arcsec$, away from the galaxy centers; these annuli contain half or slightly more of the total $K$-band flux within $20\arcsec$, so $5\arcsec$ is analogous to $r_e$ in our fitting of the quasars. For comparison with typical galaxies (i.e. without significant far-infrared excesses), we have taken typical $J-K$ colors in local bulges from \citet{Frogel1978}, and disk colors from \citet{Terndrup1994}. To guard against cases with exceptional reddening or measurement errors, we quote the median values from these two references. Table \ref{tab:extranuc} summarizes the results; the extranuclear colors of these LIRGs and ULIRGs are indistinguishable from those of normal, non-infrared-excess galaxies.  That is, despite the extreme levels of star formation in their cores, there is no evidence for significant extinction in the outer parts of the surrounding galaxies. 

This behavior is consistent with the theoretical predictions at the redshifts of our quasars by \citet{Lapi2018}, which predict that the gas, dust, and star forming  regions are all centrally concentrated in high redshift extreme star forming galaxies. 

In summary, our view of the central regions of these galaxies is obliterated by the quasars, so by necessity we are determining their masses (or upper limits to the masses) from the outskirts of the galaxy (regions outside $\sim r_e$). We therefore expect little extinction in the regions we are using to constrain the PSF normalization and therefore the galaxy fluxes and masses, and conclude that the galaxies from \cite{Mclure2011,Curtis2013} and \cite{Jiang2016} used to construct Equation \ref{eq:masses} are a reasonable comparison to our quasar hosts even in these three cases. 

\begin{deluxetable}{lccc}
\tabletypesize{\footnotesize}
\tablecaption{Summary of Observations of Outer Zones of Local ULIRGs} \label{tab:extranuc}
\tablewidth{0pt}
\tablehead{
\colhead {galaxy} &
\colhead {J-K} &
\colhead {error} &
\colhead {ref}    
}
\startdata
bulge    &  0.94 &  --&\citet{Frogel1978}  \\
disk    &  0.90 &  --&\citet{Terndrup1994}  \\
UGC 1315    &  0.98 &  0.09 &\citet{Smith1996}  \\
UGC 4509    &  0.75 &  0.09 &\citet{Smith1996}  \\
UGC 8387    &  0.79 &  0.09 &\citet{Smith1996}  \\
UGC 8696   &  0.85 &  0.09 &\citet{Smith1996}  \\
UGC 9913 (Arp 220)   &  0.90 &  0.09 &\citet{Smith1996}  \\
UGC 12332 (NGC 7469)   &  1.01 &  0.09 &\citet{Smith1996}  \\
\enddata
\end{deluxetable}
   
\subsection{Potential bias in quasar selections}

\begin{figure*}
    \centering
    \includegraphics[width=14cm]{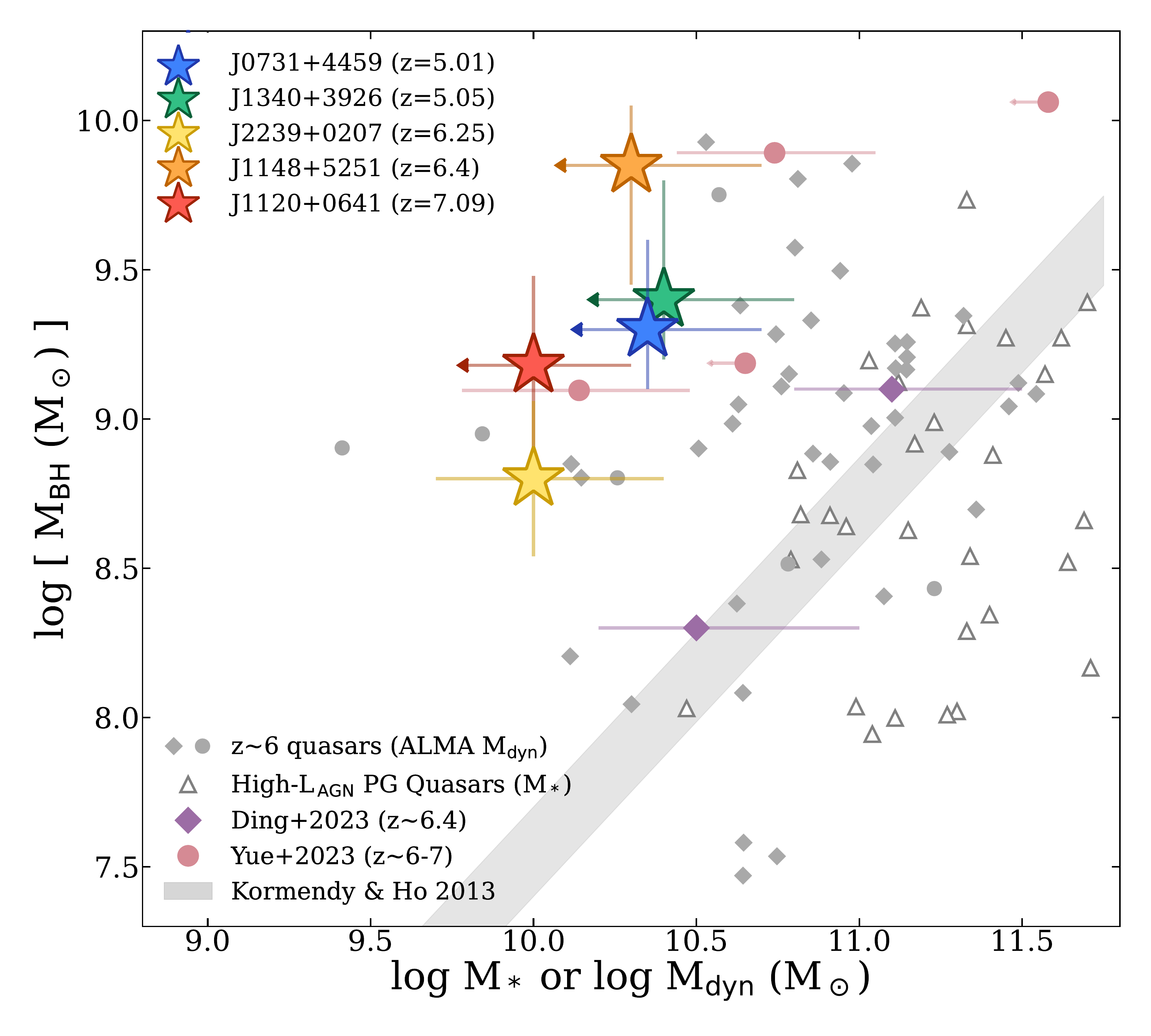}
    \caption{The \magorrian relation for our five observed quasars (stars), colored in order of increasing redshift. Black hole masses for J0731+4459, J1340+3926, and J2239+0207 are calculated from our NIRSpec spectra; masses for J1148+5251 and J1120+0641 are from the literature \citep[][respectively]{Barth2003}. For reference, we also show the two $z\sim6.4$ quasars whose host galaxies were detected via PSF subtraction by \cite{Ding2023} (purple diamonds), and the six $z\sim6-7$ quasars from \cite{Yue2023}. The local \magorrian relation in local non-AGN systems from \cite{Kormendy2013} (grey shaded region) is shown. $z\sim6$ quasars with ALMA dynamical masses \citep{Izumi2021,Pensabene2020} are shown as grey diamonds and circles respectively; their distribution is offset from that of local high-$L_{\mathrm{AGN}}$ quasars (white triangles), demonstrating that the high-$z$ offset from the local \magorrian relation is not simply a consequence of selection biases. The constraints on the host galaxy masses of our five $z\sim5-7$ quasars and the EIGER sample \citep{Yue2023} put them all well above the local \magorrian relation, while the \cite{Ding2023} galaxies appear to be consistent with the local relation. This could point to a greater spread in the ratio $M_{\mathrm{BH}}/M_*$ at high redshift than is observed locally.}
    \label{fig:magorrian}
\end{figure*}

Even with our conservative upper mass limits, all five of our quasars lie significantly above the local \magorrian relation: the central SMBHs are overmassive compared to their host galaxies. Similar behavior is found for the EIGER quasar sample, whose members also lie above the local \magorrian relation \citep[][pink circles in Figure \ref{fig:magorrian}, though they have significantly larger radii for their host galaxies than we find for ours]{Yue2023}. In contrast, the two quasars whose host galaxy emission was detected via subtraction of an empirical PSF by \cite{Ding2023} (purple diamonds in Figure \ref{fig:magorrian}) have host galaxies that are, within their uncertainties, roughly as massive as predicted from their black hole mass by the local \magorrian relation of \cite{Kormendy2013}. 

Overall, these results strongly suggest that the SMBHs in the high redshift quasars are overmassive compared with their host galaxies. However, it is critically important to consider the selection effects at play when drawing conclusions from high-redshift quasar samples. In particular, our sample of luminous $z\sim6$ quasars and hosts is subject to two significant biases.

The first is simple: more massive host galaxies are significantly easier to detect and constrain after performing PSF subtraction. If $z\sim6$ quasars of a given black hole mass display a range of host galaxy masses, the more massive hosts (i.e., those galaxies lying to the right of the \magorrian relation) are more likely to be detected. This bias, however, is contrary to our results. No $z\sim6$ quasar host galaxies detected with JWST to date lie to the right of the local relation (as can be seen in Figure \ref{fig:magorrian}).

The situation is complicated, however, by the selection bias first described by \cite{Lauer2007}. Due to the steep slope at the bright end of the galaxy luminosity function, the most massive black holes are more likely to lie in lower-mass galaxies (where the black holes are overmassive compared to the local \magorrian relation), than in high-mass galaxies where the massive black hole is proportional. Because massive black holes can accrete at higher rates and produce brighter AGN, this bias may then also manifest in flux-limited quasar samples. Therefore, to accurately compare our sample to local AGN and draw conclusions about the \magorrian relation at high redshift, we must use a similarly biased sample of the most luminous quasars in the local Universe.

As discussed in more detail in S23, we create this comparison $z\sim0$ sample from the Palomar Green (PG) sample of Type 1 quasars at $z\leq0.5$. This provides a valid comparison because the high redshift quasars are remarkably similar in all the relevant properties to very luminous local type-1 quasars \citep{Fan2023}.  By examining PG quasars with $L_{\mathrm{AGN}} \geq 2\times10^{12} L_\odot$ (shown in Figure \ref{fig:magorrian} as open triangles), we recreate the same luminosity bias present in our high-$z$ sample and others in the literature \citep[e.g.][]{Ding2023, Yue2023}. The stellar and black hole masses we retrieve for the PG quasars are derived using methods as similar as possible to those used for our $z\sim6$ quasar sample: stellar masses from rest-frame optical photometry and black hole masses from single-epoch spectra (see S23 for more details). If these $z=0$ luminous quasars behaved like the $z\sim6$ JWST samples, there would be a significant population lying to the left of the local relation: this is not the case. The local quasars lie on or to the right of the local relation, with only a very small minority of the sample (four quasars out of 37 total) slightly to the left. Indeed, a two-sample KS test comparing the distance above/below the local \magorrian relation for the high-$L_{\mathrm{AGN}}$ PG quasars and the $z\sim6$ samples of this work, \cite{Ding2023}, and \cite{Yue2023} returns a p-value of $4\times10^{-6}$, which indicates these samples are drawn from distinct populations (despite both being subject to the same luminosity bias). The \cite{Lauer2007} bias therefore cannot be entirely responsible for the increasing number of high-redshift quasars far from the local \magorrian relation.

\subsection{Implications for Co-Evolution of SMBHs and Host Galaxies}

It is therefore unlikely that the position of our quasars above the local \magorrian relation is only a consequence of extinction effects on our estimates of the host masses, or of selection biases. Instead, we find that even the most massive SMBHs at $z \sim 6$ share the behavior found for lower mass ones \citep[e.g.,][]{Harikane2023,Goulding2023, Maiolino2023, Kokorev2023}, namely to show a large scatter in $\rm{M_{BH}/M_*}$ with a strong tendency toward over-massive SMBHs compared with the local \magorrian relation. The small number of SMBHs discovered so far at even higher redshifts \citep{Bogdan2023, Goulding2023, Kokorev2023, Pacucci2023, Ubler2023} share this behavior.

The spread of \magorrian ratios observed---from galaxies consistent with the local relation to cases an order of magnitude less massive than predicted---aligns with existing measurements of [C~II] dynamical masses using ALMA and other sub-mm telescopes (see the small grey diamonds in Figure \ref{fig:magorrian}). Indeed, these ALMA results eliminate the possibility that the local \magorrian relation holds for {\em gas} masses rather than stellar masses at $z\sim6$: ALMA dynamical masses at this redshift also tend to indicate galaxies less massive than predicted by the local \magorrian relation. However, because the gas dynamics can be influenced by non-gravitational forces and the mass estimates are dependent on assumptions about the gas distribution around the SMBH, these results should be interpreted with caution until stellar masses can be determined \citep{Inayoshi2020}. An illustrative case is the host galaxy of J2239+0207, with a stellar mass more than an order of magnitude less than its dynamical mass (S23). Such galaxies are not unheard of in the high-redshift Universe \citep[see e.g.][for an example of a $z=6.34$ non-active galaxy with stellar mass nearly an order of magnitude less than its dynamical mass]{Riechers2013}. 

Clearly, not all galaxies---or even all AGN---at $z\sim6$ will lie significantly above the local relation: even some luminous quasars with detected host galaxies are consistent with it. However, the number of luminous quasars observed to have overmassive black holes relative to the stellar masses of their host galaxies (a rarity at z=0), taken together with similar results from sub-millimeter telescopes, implies that there is additional scatter in the \magorrian relation at high-redshift: the relationship between galaxy mass and black hole mass is not nearly as tight as in the local Universe. One Gyr after the Big Bang, supermassive black holes and their host galaxies do not yet march in lockstep.

These results appear to contradict predictions that the growth of galaxy mass will precede that of SMBH mass such as in \citet{Li2008}. They are also problematic for the some of the simulations summarized by \citet{Habouzit2022}, which predict reasonably coordinated co-evolution. These results suggest that the growth of the SMBHs is not purely due to galaxy mergers, where the overall host galaxy mass grows and much of the interstellar gas falls to the center where it can be accreted by the SMBH. Nonetheless, the presence of metals and dust around these quasars \citep[e.g.,][]{Maiolino2003, Venemans2018} shows that they must lie in evolved populations of stars.  

The mystery of the rapid SMBH growth therefore persists \citep{Fan2023}. Solutions may well require ways that the growth of SMBHs could be turbocharged to proceed much more rapidly than in the conventional major merger scenario, as discussed for example by \citet{Inayoshi2020}.

\section{Summary and Conclusions} \label{sec:summary}

We obtained NIRCam images and NIRSpec IFU spectra of a sample of high-luminosity $z\sim5-7$ quasars, and perform careful point-spread function subtraction to detect emission from their host galaxies. 

\begin{itemize}
    \item Utilizing NIRCam medium bands and reference PSFs from dedicated reference star observations provides excellent quasar-star PSF agreement, allowing us to probe to within less than a kiloparsec of the galaxies' centers before PSF artifacts dominate.
    \item We place lower and upper limits on the rest-frame optical fluxes of the host galaxies by iterating the normalization of the reference PSF, allowing us to precisely determine the maximum flux attributable to the host rather than the quasar. By relating the F360M flux to galaxy mass via Spitzer/IRAC Band 1 observations of galaxies from $5<z<8$ in the literature, we find that our entire sample lies significantly above the local \magorrian relation, with mass upper limits $<10^{11} M_{\odot}$.
    \item We explore the possibility that our host galaxies are highly extincted and therefore appear to be less luminous than expected: based on the metallicities and extinctions of typical $z\sim6$ galaxies and the dust morphology of local extremely star-forming galaxies analogous to the highly star-forming galaxies in our sample, we conclude that extinction at the radii we are able to probe with PSF subtraction is likely not significant. We also examine the possibility of selection biases driving the apparent deviation from the local \magorrian relation observed at $z\sim6$, and conclude that even bias towards finding overmassive black holes associated with luminous quasars cannot explain the abundance of overmassive SMBHs observed at high redshift.
    \item Analysis of mm-wave gas dynamics measured for low-redshift PG quasars indicates that selection biases also do not influence these results sufficiently to account for the behavior of the high-z quasar hosts.
    \item Our results appear to contradict evolutionary models where the SMBH growth at very high redshift proceeds much as it does locally, through the mergers of galaxies and injection of gas into the nucleus where it is accreted by the SMBH, with co-evolution of the SMBH and its host galaxy linked from early times.
\end{itemize}

JWST's NIRCam and NIRSpec instruments are rapidly expanding the number of quasar host galaxies detected at $z\gtrsim4$, demonstrating that the relationship between SMBH and galaxy masses displays much more scatter at high-redshift than locally.  
As the theory of SMBH-galaxy coevolution evolves, such high-redshift luminous quasars will continue to provide a challenge for models to explain the apparent disconnect between SMBH and galaxy masses at $z\sim6$ now being uncovered with JWST.

\begin{acknowledgments}
We thank Junyao Li for bringing a calculation error to our attention, and the referee for helpful comments that improved the clarity of this paper. MS, SA, JL, GR, and KH acknowledge support from the JWST Mid-Infrared Instrument (MIRI) grant 80NSSC18K0555, and the NIRCam science support contract NAS5-02105, both from NASA Goddard Space Flight Center to the University of Arizona. The JWST data presented in this paper were obtained from the Mikulski Archive for Space Telescopes (MAST) at the Space Telescope Science Institute. The observations can be accessed via\dataset[DOI.]{http://dx.doi.org/10.17909/4j9v-9q21}
\end{acknowledgments}

\vspace{5mm}
\facilities{JWST(NIRCam, NIRSpec)}

\software{Astropy \citep{Astropy2013,Astropy2018},  
         Matplotlib \citep{Hunter2007}, NumPy \citep{VanDerWalt2011}, photutils \citep{Bradley2022},
         }

\bibliography{bibliography}{}
\bibliographystyle{aasjournal}

\end{document}